\documentclass[seceq,supplement]{ptptex}

\usepackage{graphicx}

\title{Puzzles with Tachyon in SSFT
                                    and
                 Cosmological  Applications }

\author{Irina \textsc{ Aref'eva}}

\inst{Steklov Mathematical Institute, Russian Academy of Sciences}

\abst{
This work is my contribution to the proceedings of the conference
``SFT2010 -- the Third International Conference on String Field Theory and Related
Topics''. We discuss  properties of nonlocal tachyons and
 nonlocal SFT inspired cosmology.
}

\newcommand{\be}{\begin{equation}}
\newcommand{\ee}{\end{equation}}
\newcommand{\bea}{\begin{eqnarray}}
\newcommand{\eea}{\end{eqnarray}}

\begin{document}

\maketitle

\section{Introduction}

In this talk I would like to discuss time dependent solutions for the tachyon field in
cubic (super)string field theories. The key new property of the tachyon field is that
it satisfies  a
nonlocal equation.

The plan of this talk is the following:
\begin{itemize}
\item Yukawa  field and string field
\item Nonlocality in SFT and SFT inspired models
\item Applications of SFT nonlocality to cosmology
\item Mathematical questions around nonlocal cosmology
 \end{itemize}

 \subsection{Yukawa  field and string field}
 We are at the Yukawa Institute and
it is worth to remind that in 1949 Yukawa to remove divergences in
QFT have proposed a nonlocal model, which  is free from the
restriction that field quantities are always point like functions in
the ordinary space \cite{Yukawa}. The Yukawa field $U(x,r)$ depends
on the spacetime coordinates $x$ and an extra vector  $r$ and
is the subject of the restriction \be (r_\mu r^\mu-\lambda^2)U(x,r)=0,
\ee as well as a solution
to the Klein-Gordon equation. The main motivation to deal with nonlocal
models in old days was the elimination of divergences, see also
\cite{DIB,DAK,GM,PU}. There is a similarity between the expansion
\be
U(x,r)=\phi(x)+A_\mu (x)r^\mu +B_{\mu \nu}(x)r^\mu r^\nu+...
\ee
 and
the SFT expansion
\be
\Phi[x(\sigma)]=\phi(x) +A_\mu
(x)\alpha^\mu_{-1}+...\,.
\ee
One can say that
the Yukawa  field $U(x,r)$ is  a prototype of the modern
string field $\Phi[x(\sigma)]$.

\subsection{Nonlocality in strings}
Elimination  divergences is also one of the main goals of string
theories. Generally speaking,  nonlocality in string theories  related with
extended character of the  Nambu-Goto string and is related with the string parameter  $\alpha^\prime$.

\subsection{Nonlocality in String Field Theory}
\subsubsection{In fundamental setting}

Nonlocality is a characteristic feature of noncommutative geometry.
The covariant Witten string field theory (SFT)\cite{Witten} is an
example of noncommutative geometry. The light-cone Kaku-Kikkawa  SFT
\cite{KK} and covariant light-cone-like HIKKO
(Hata-Itoh-Kugo-Kunitomo-Ogawa)  \cite{HIKKO} SFT have nonlocal
features. The same concerns the cubic fermionic AMZ-PTY (Aref'eva,
Medvedev, Zubarev and Preitschopf, Thorn, Yost) SFT
\cite{AMZ-PTY} as well as the fermionic nonpolynomial Berkovits SFT
\cite{Nat}. Nonlocality in
SFT is a subject of several discussions \cite{EWood,Hata,Gross}.

\subsubsection{In practical setting}
By practices I mean cosmological applications. In cosmology we study time-dependent
solutions in a
nontrivial cosmological metric,
\be
ds^2=-dt^2+a^2(t)dx^2.
\ee
 We do not have in our disposal
 SFT in a nontrivial cosmological background. We start from a discussion of
time-dependent
solutions in the flat case.
The equation of motion for the cubic SFT has a simple form
\be
\label{EOM}
Q\Phi+\Phi\star\Phi=0\ee
Because the BRST operator $Q$ has nontrivial cohomologies, the zero-curvature equation
(\ref{EOM})  has nontrivial
solutions. An active search for  Higgs-type solutions of (\ref{EOM}), i.e. solutions
for which all component fields are constant,
  began in 1999. This activity was
primarily associated with the Sen conjectures (see \cite{review-SFT}).
There were many attempts to find numerical solutions
of  (\ref{EOM}) and verify the Sen conjectures in the framework
of the level-truncation method.
The impulse for a new development came from a  remarkable Schnabl's paper \cite{MS}.
In this paper explicit solutions of Eq. (\ref{EOM}) have been found
for the boson string (see refs.\cite{Okawa,ES} for simplifications of this solution).
  In subsequent papers \cite{ET1,AGM,AGKMM,RG,Arroyo:2010fq,AG} solutions have been found
 for the NS fermionic string. In the bosonic case  level-truncated vacuum solutions
 are rather closed to the exact one. In fermionic case the analytical solution
\cite{AGM}
starts from the level-truncated solution \cite{ABKM}. There are also exotic universal solutions \cite{ET2}.

 There were several attempts to find  time depending
  level-truncated solutions \cite{MZ,FH,yar,AJK,NM} as well as exact solutions. There is a hope that study of marginal deformed theory would help to
 find exact rolling solutions \cite{Schnabl:2007az,tds,Kiermaier:2007ba,Kwon:2008ap,Kawano:2008jv,Kiermaier:2010cf,Bonora}.
 However
finding analytical time depending solutions suitable for
cosmological applications is still an open problem in SFT.
By this reason and motivated that leading level-truncated versions of SFT actions
reproduce qualitatively known exact results, we discuss cosmological applications
within a level-truncated model
\cite{IA}.

\section{Nonlocal Cosmology from String Field Theory}
\subsection{Model}

Our nonlocal cosmological model is given by the following action \cite{IA}
\be
\label{NLCM}
S=\int
d^4x\sqrt{-g}\left\{\frac{R}{2\kappa^2}+\frac1{\lambda_4^2}\left(-\frac{\xi^2\alpha^\prime}{2}
g^{\mu\nu}\partial_{\mu}\phi(x)\partial_{\nu}\phi(x)+\frac1{2}\phi^2(x)-
\frac{1}{4}\Phi^4(x)-T^\prime\right)\right\}
\ee
Here
$g_{\mu\nu}$ is the four-dimensional metric,  $\kappa$
is the gravitational constant and
$\lambda_4$ is the scalar field coupling constant.
  In this SFT nonlocal model our Universe is considered as a D3
non-BPS brane embedded in the 10-dimensional space-time.
Due to this embedding we have
\be
\frac1{\lambda_4^2}=\frac{v_6 M_s^4}{g_o}\left(\frac{M_s}{M_c}\right)^6,
\ee
$g_o$ is
the open string dimensionless coupling constant, $M_s$ is the string
scale $M_s=1/\sqrt{\alpha^\prime}$ and
 $M_c$ is a scale of the compactification,  $v_6$ is a number related with
 a volume of
 the  6-dimensional compact space.
  The role of
the dark energy plays the Neveu-Schwarz (NS) string tachyon leaving in the GSO$-$
sector. The tachyon action is dictated by  the cubic fermionic SFT \cite{ABKM}
 and it is  nonlocal due to string   effects.
 The form of the nonlocal interaction defined by the cubic
fermionic SFT (CFSFT) \cite{ABKM} is in fact
a more complicated as compare with $\Phi^4$,
 where $\Phi$ is related with
the tachyon field $\phi$ by the following relation
\be
 \label{FR}
\Phi=e^{\frac{\alpha^\prime}{8}\Box_g}\phi, ~~~{\mbox {where}}~~~
\Box_g=\frac1{\sqrt{-g}}\partial_{\mu}
\sqrt{-g}g^{\mu\nu}\partial_{\nu}.\ee
But, by analogy with
the flat case \cite{yar,AJK} we believe that
the approximation accepted in (\ref{NLCM})
reflects essential physical properties of the model.
 $\xi^2\approx 0.9556$ is a constant defined by the CFSFT.

The potential has perturbative and nonperturbative minima.
A transition from a perturbative vacuum to a non-perturbative one  is interpreted as
D-brane decay. In the flat background the
D-brane tension  $T$  is
equal to 1/4, and this value is compensated by the minimal value of the tachyon
potential (Sen's conjecture), so that the total vacuum energy is zero.
This compensation means that the cosmological constant is zero,
\be\Lambda \equiv T+V_0=0.
\ee

In (\ref{NLCM}) we  postulate  a minimal form of the tachyon interaction with gravity.
The total energy of the system in the true non-perturbative vacuum
we interpret as the cosmological constant,
  \be
 \Lambda ^\prime\equiv T^\prime+V_0.
  \ee
  It has been conjectured \cite{IA1}  that an existence
of a rolling solution describing a smooth transition to the true vacuum
in a given cosmological background does define the value of the cosmological
constant.
We cannot prove this conjecture but
arguments to its favor were given using the local approximation \cite{AKV}.
A recent breakthrough in solving numerically the full nonlinear and nonlocal  system
of equations \cite{LJ}
also supports this conjecture.

Under some
conditions our nonlocal  model admits a local approximation and
displays  a phantom behaviour \cite{AJK}. Note that
unlike phenomenological phantom models here the phantom appears in an effective
theory. Since SFT  is a consistent theory this approach does not suffer from usual problems
which are inevitable for phenomenological phantom models.

 \subsection{Problems}
The questions that  we  address in nonlocal cosmology are concerned  two different
epochs
of the Universe evolution, namely physics just after  Big Bang and the modern
evolution. \begin{itemize}\item About the modern evolution epoch we would like to know
\begin{itemize}
\item why now  the cosmological constant is  so small;
\item can we construct a physically acceptable dynamical dark energy (DE)
model with  $w < - 1$
\item
                can we get a periodic crossing the $w=-1$ barrier
  \end{itemize}
  It is natural to rise these questions since present cosmological observations
               do not exclude an evolving  DE and according to recent data
we have for the state parameter $w=-1.04\pm 0.06.$ Local
DE models with the state parameter $w < -1$ violate the
null energy condition (NEC).
\item As to the early time evolution we would like to know
\begin{itemize}\item
can we construct an inflation nonlocal model with a large non-Gausianity;
\item
can we estimate influence of presence of nonlocal matter on  primordial black holes
formation

\end{itemize}
About applications of  nonlocal SFT models to the DE problem see
\cite{IA,IA1,AKV,AK,LJ}, to inflation see
\cite{Lidsey,Cline,MN:2008_AIP,MN:2008,Barnaby:2008vs} and to
cosmological singularity see refs. in \cite{AJ}.

\end{itemize}

\subsection{How we study the model and what we get}
In the spatially flat FRW metric
the dynamics in the model (\ref{NLCM}) is described by a system of two
nonlinear nonlocal equations \cite{IA} for the
tachyon field
 and the Hubble parameter $H(t)=\dot{a}/a$
\begin{eqnarray}
\label{EOM_ST0approx_phi}
 \left(\xi^2{\cal D}+1\right)e^{-\frac{1}{4}{\cal D}}\Phi &=&\Phi ^3,
\\
\label{EOM_ST0approx}
3H^2&=&\frac{\kappa^2}{\lambda_4^2}\left(\frac{\xi^2}{2}\partial_t
\phi^2-\frac{1}{2}\phi^2+\frac{1}{4}\Phi^4+{\cal E}_1+{\cal E}_2+T^\prime\right),
\eea
where ${\cal D}=-\partial _t^2-3H(t)\partial_t,\,\,\,\,\,H=\partial_t a/a$ and
\bea
{\cal E}_{1}&=& - \frac{1}{8}\int_0^1 ds\left((\xi^2{\cal D}+1)\,\,\,
e^{\frac{s-2}{8}{\cal D}}
\Phi \,\right)\cdot
\left({\cal D}\,\,
e^{-\frac{1}{8} s {\cal D}} \Phi\right),\\
{\cal E}_2&=& -\frac{1}{8} \int_0^1 ds\left(
\partial_{t}(\xi^2{\cal D}+1)\,e^{\frac{s-2}{8}{\cal D}}
\Phi \right)\cdot
\left(\partial_t
e^{-\frac{1}{8}s{\cal D}} \Phi\right).
\end{eqnarray}
The nonlocal energy ${\cal E}_{1}$
plays the role of an extra potential term and ${\cal E}_{2}$ the role of the
kinetic term. Note that here we use a dimensionless time $t\to t \sqrt{\alpha^\prime}$.

Equations (\ref{EOM_ST0approx_phi}) and (\ref{EOM_ST0approx}) form a rather
complicated system of  nonlinear nonlocal equations for functions $\Phi$ and $H(t)$ because of the presence of an infinite
number of derivatives and a non-flat metric.
Before to discuss  the methods of study  this model let us mention the known
methods of study  equation (\ref{EOM_ST0approx_phi})
in the flat background, $H=0$,
\begin{equation}
\left(-\xi^2\partial_t ^2+1\right)e^{\frac{1}{4}\partial_t ^2}\Phi(t)=\Phi(t)^3.
\label{EOM_ST0approx_flat}
\end{equation}
$\xi=0$ corresponds to p-adic string (see \cite{padicrev}and refs therein). Equation (\ref{EOM_ST0approx}) in the flat case describes the energy conservation \cite{AJK}.
A boundary
problem $\Phi(\pm \infty)=\pm 1$ for (\ref{EOM_ST0approx_flat}) has been studied
using different methods.  A numerical method \cite{yar}
is based on an integral representation of (\ref{EOM_ST0approx_flat}).
 The integral  representation
 has also been used to prove existence theorems \cite{VV,Jouk,Prokh}
and  is also
 related with a diffusion equation method \cite{VS}. This method
 uses an auxiliary function of two variables $\Psi(\lambda,t)$
satisfying  the diffusion equation
\be
(\partial _\lambda-\partial_t ^2)\Psi(t,\lambda)=0,\ee
$\Phi(t)= \Psi(t,0)$ and Eq.(\ref{EOM_ST0approx_flat}) becomes the boundary condition
\be
\left(-\xi^2\partial_t ^2+1\right)\Psi(t,\frac14)=\Psi(t,0)^3\ee

 The following  characteristic  properties of  (\ref{EOM_ST0approx_flat})
 have been obtained. First is
an existence of a  critical point $\xi^2_{\text{cr}}\approx 1.38$ such that for
$\xi^2<\xi^2_{\text{cr}}$
eq. (\ref{EOM_ST0approx_flat}) has  a rolling solution \cite{yar} interpolating between
$\pm 1$. Second is
 an existence of a dominance of an extra nonlocal
kinetic term  ${\cal E}_{2}$  over  the local kinetic one \cite{AJK}
and as a result, an appearance  of  a phantom behavior providing $w<-1$.

There is also a method of  decomposition on local fields \cite{AV,AK,AJV,Nardelli}.
This method works well for linear equations
and has been used to study solutions to
(\ref{EOM_ST0approx_flat}) near vacuum $\pm 1$.
It is very interesting to   find  approximate models admitting
explicit solutions and having above mentioned properties.
They  could contain  two or more components local fields.
Let us also mentioned
  almost exact  solutions methods \cite{Forini,AJ}.

An investigation of    non-flat eqs.
  (\ref{EOM_ST0approx_phi}) and (\ref{EOM_ST0approx}) is essentially more complicated.
A numerical study has been performed in \cite{LJ}.
A decomposition on local fields  have been
used in \cite{AKV,AJ,AKV2,AJV}.
A simplest one phantom mode approximation with  a special six-order potential
\cite{AKV}
has  the
solution
\be
\phi(t)=\tanh (t)
\ee
 and gives $\Lambda\equiv 1/4+V_0^\prime=1/m^4_p$,
 where $m^4_p$ is the reduced Planck mass
 \be
 m^2_p\sim \frac{M_s^2}{M_p^2}(\frac{M_s}{M_c})^6. \ee
 The difference $V_0^\prime-1/4$ is assumed  coming
 from nontrivial
 background effects. Assuming also that
$M_c\sim M_p$ and $M_s\sim 10^{-6.6}M_p$
we get a possibility to explain the small value of
the Hubble parameter in the real time $H_0\sim M_s
/m_p^2$, namely
$$H_0\sim 10^{-60}M_p.$$

 \section{Generalization of SFT cosmology}
\subsection{Models}
A generalization of the nonlocal SFT cosmological model (\ref{NLCM}) has the form
\be
\label{GNLCM}
S=\int
d^4x\sqrt{-g}\left\{\frac{R}{2\kappa^2}+\frac{1}{g^2_o}\left(\frac{1}{2}
\Phi(x)F(\Box_g)\Phi(x)-V(\Phi)-T^\prime\right)\right\},
\ee
where $F(z)$  in a  neighborhood of the point
$z=0$ is  an analytic function and
\begin{equation}
\label{Fser}
F(\Box_g)=\sum\limits_{n=0}^{\infty}f_n\Box_g^{\;n}.
\end{equation}
Such type of models have been considered in \cite{AV,AJV,AK}.
A special interest presents the case of  $\zeta (\Box)$, where $\zeta(s)$
is the famous Riemann zeta-function \cite{AI-IV-Riem}. One of
the reasons  to consider this case is that the Riemann zeta-function
 is  universal in the sense that any
analytic function can be approximated by the shifts of the Riemann zeta-function.

Modified gravity cosmological models have been proposed in the hope of
finding to solutions to the open problems of the standard cosmological
model. There are a lot of ways to deviate from the
Einstein gravity.  Different modifications of gravity are  considering nowadays
in the literature.
A simple nonlocal gravity has the form
\begin{equation}
\label{S3} S_3=\int d^4 x \sqrt{-g}\left\{ \frac{1}{{16\pi
G_N}}R\left(1 + {\cal F}( L^2\Box )R\right) + {\cal L}_{\rm matter}
\right\},
\end{equation}
where ${\cal F}(z)$ is  an analytic function at the point $z=0$.

\subsection{Mathematical questions around nonlocal cosmology}
\subsubsection{Nonlocal kinetic operator as a pseudo-differential operator}
Let us consider the nonlocal Klein--Gordon equation
\be F(\Box)\phi=0, \label{no-KG} \ee
If $F(z)$ is a polynomial, then (\ref{no-KG}) is a  partial differential equation,
otherwise (\ref{no-KG}) is a  pseudo-differential equation
and  we have to pay a special attention to a definition of
operator $F(\Box)$.
In the flat space-time, $(t,x_i)\in M^4$,
$\Box={}-\partial^2_t+\partial_{x_i}\partial_{x_i}$ and $F(\Box)$ is
defined via the Fourier transform \cite{PU,MZ,VV,AV}
\be
\label{ft-sym} F(\Box)\phi
(x)=\frac{1}{(2\pi)^4}\int\,F(-k^2)\tilde\phi(k)e^{-ixk}d^4k,\,\,\,k^2=-k_0^2+k^2_i,\qquad i=1,2,3.
\end{equation}
where $\tilde \phi (k)$ is a Fourier transform of a function $\phi(x)$,
$\tilde \phi(k)=\int\phi(x)e^{ik
x}d^4x
$.

In cosmology we usually deal with a positive time variable, $t>0$,
and it is more suitable to use
 the Laplace transform of the time variable
 and the Fourier transform \footnote{We do not make difference in notations for
 the Fourier
 and Laplace transforms, supposing that the meaning is clear from the context.}
 at space variables \cite{BK:2007}, $\tilde
\varphi(s,\vec{k})=\int\limits_{0}^\infty dt\int d\vec{x}\varphi(t,\vec{x})
e^{-st-i\vec{k}\vec{x}}
$. The inverse transform is defined as
$\varphi (t,\vec{x})=(2\pi i)^{-4}
\,\int\limits_{c-i\infty}^{c+i\infty} d\!s\int d\vec{k}
\tilde\varphi(s,\vec{k})e^{ts+i\vec{k}\vec{x}}.
$

For an analytic function $F(z)$ at $z=0$  we
 understand the operator $F(\Box)$
as
\bea
\label{pseudo_app}
  F(\Box)\varphi(t,\vec{x}) &=& \frac{1}{(2\pi)^4 i}\int\limits_{c-i\infty}^{c+i\infty}d\!s
  \int d\vec{k}\, e^{st+i\vec{k}\vec{x}} \left\{ F(-s^2-\vec{k}^2)
  \tilde{\varphi}(s,\vec{k})+r(s,\vec{k}) \right\},\\r(s,\vec{k})&=&- \sum_{n=1}^{\infty} \sum_{j=1}^n
  \frac{s^{2n-j}}{n!}\left.\left[\frac{\partial^n}{\partial w^n}F(-w-\vec{k}^2)\right]
  \right|_{w=0}\varphi^{(j-1)}(0,\vec{k}).
\eea
This definition is based on the following definition of the
time derivatives
\bea
\label{app_der}
  \partial_t^{n}\phi(t) = \frac{1}{2\pi i}\int\limits_{c-i\infty}^{c+i\infty} d\!s\, e^{st}
  s^n \left[ \tilde{\phi}(s) - \sum_{j=0}^{n-1 }\frac{\partial_t^{(j)}\phi(0) }{s^{j+1}}  \right]
\eea
here
$\tilde
\phi(s)=\int\limits_{0}^\infty\phi(t)e^{-st}dt$.
Definition
(\ref{app_der}) has meaning under an assumption that the function
$
{\ae}_n(z)=z^n\left[ \tilde{\phi}(z) - \sum_{j=0}^{n-1}z^{-j-1}
\partial_t^{(j)}\phi(0)  \right] $
is holomorphic at
$\Re \,z \geqslant a$ and satisfies the following condition
$|{\ae}_n(z)|={\cal
O}(|z|^{-1-\alpha}),\,\,\,\mbox{for} \,\,\,|z|\to\infty,\,\,\,\Re\, z \geqslant a
\,\,\mbox{and }\,\,a\leqslant c.
$
Definition (\ref{app_der}) is in agreement with the property
\be
\label{mul-der}
 \partial_t^{m} \partial_t^{n}\phi(t) =\partial_t^{n+m}\phi(t).
\ee

\subsubsection{The Weierstrass product and factorization of nonlocal kinetic term}
To study the nonlocal Klein--Gordon equation (\ref{no-KG}) we use
\cite{AV,AJV,AI-IV-Riem} (see \cite{PU} for early used of the
Weierstrass product)
the Weierstrass factorization theorem.
According to the theorem for any entire function $F$ which is
not identically zero, has  $m$-order zero
at
$0$, and has the non-zero counting multiplicity zeroes  $\{z_k\}$
 there exist non-negative integers $p_1,
p_2,...$ and an entire function $Q_0(z)$
 such that:
\begin{equation}
F(z)=z^me^{Q_0(z)}\prod_{k=1}^\infty\left(1-\frac{z}{z_k}\right)e^{Q_{k}(z)},\,\,\,\,\,Q_{k}(z)=\sum_{l=1}^{p_k}
 \frac1l\left(\frac{z}{z_k}\right)^l,
\end{equation}

Let us consider equation \begin{equation}
 \label{WP-short-box-eq}
F(\Box)\phi=-J,\,\,\,\,
\end{equation}
for function $F$ that has
a finite number of zeros,
$ F(z)=e^{\mathfrak{f}(z)} \prod\left(z
-z_k\right),$
and
$\mathfrak{f}(z)=\lambda z, \lambda>0$.
If we understand the differential
operator via its Fourier or Laplace transforms ((\ref{ft-sym}), (\ref{pseudo_app}))
 then from the factorization
of the function $F(z)$ follows the
factorization property of the operator $F(\Box)$
\be
\label{WP-d-box}
F(\Box)=e^{\lambda\Box } \prod\left(\Box -z_k\right),
\ee
and Eq.(\ref{WP-short-box-eq}) reduces to two problems
\be
1)\,\prod (\Box -z_{k})\psi ={}-J\,\,\,\mbox{and}
\,\,\,\,\,2)\,e^{\mathfrak{f}(\Box)}\phi=\psi,\label{tw--eq}\ee
A solution to the second equation in  (\ref{tw--eq}) depends on the definition  of the
operator
$e^{\mathfrak{f}(\Box)}$ \footnote{ This
form of nonlocal models has been considered by Pais and Uhlenbeck
\cite{PU}.
Their motivation for the choice of function $F(\Box$ )
is that in the corresponding field problem the zeros of $F(z)$ will
each correspond to quanta of certain mass and the exponential
could serve of a cut-off factor.
}.
To solve this equation for $\mathfrak{f}(\Box)=\lambda \Box$ we use
the fifth Fock's parameter  method.
In this method, one considers the function  of  two arguments
$\Psi(t,\lambda)=e^{\lambda \Box}\phi(t)$
that
solves the Fock equation (more precisely, the Euclidean version of this equation, $\lambda=i\tau$, where $\tau$
is the Fock fifth parameter~\cite{Fock}):
\be
\label{d-e}
\partial _\lambda\Psi(x,\lambda)=\Box\Psi(x,\lambda)\ee
and the boundary condition
\be
\label{bc1}
\Psi(x,\lambda^\prime )|_{\lambda^\prime=0}=\phi(x).\ee
In this framework the second equation in (\ref{tw--eq}) also can be written  as the boundary condition
\be
\label{bc2}
\Psi(x,\lambda^\prime)|_{\lambda^\prime=\lambda}=\psi(x).\ee

Therefore, a solution to the second equation in (\ref{tw--eq})
is reduced to a solution of the boundary problem for the Fock equation
\footnote{$\Box \to \,\,\partial_i\partial_i$ gives the diffusion (heat) equation.}. These formal considerations
 can be put on the mathematical background, in particular, in the case of the space homogeneous configurations.

\subsubsection{Exponential equation with quadratic derivatives via Fourier transform.}

Let us take $\mathfrak{f}(z)=\lambda z$, $\lambda>0$ and consider
\bea
\label{mech-exp}
e^{\lambda \partial
_t^2}\phi(t)=\psi(t),
\eea
$-\infty<t<\infty$,
 as  a
linear pseudo-differential equation with the symbol $e^{-\lambda
k^2}$. We immediately get the integral form of this equation
\be
\label{Phi-K}
{\cal K}[\phi](t, \lambda)=\psi(t)\ee
where
${\cal K}[\phi](t,\lambda)\equiv\frac{1}{\sqrt{4\pi \lambda}}\int\limits_{-\infty}^\infty \,\phi(t')\,
e^{-\frac{(t-t')^2}{4\lambda}}\,dt'.
$
Denoting $\Psi(t,\lambda)\equiv e^{\lambda \partial
_t^2}\phi(t)$ we get
\be
\partial_\lambda \Psi(t,\lambda)=\partial^2_t\Psi(t,\lambda),\,\,\,\,\,\Psi(t,\lambda)|_{\lambda=0}=\phi(t)
,\,\,\,\,\,\Psi(t,\lambda^\prime)|_{\lambda^\prime=\lambda}=\psi(t)
\ee

\begin{figure}[h!]
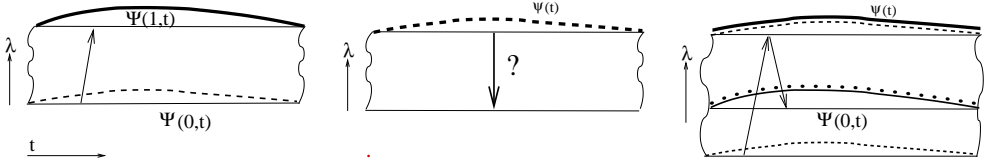

\begin{center}
\includegraphics[width=4cm]{bc-bc-color.eps}$\,\,\,\,\,$
\includegraphics[width=4cm]{bc-bc-inv-color.eps}$\,\,\,\,\,$
\includegraphics[width=4cm]{bc-bc-inv-a-color.eps}$\,$
\end{center}
\caption{\label{bc-bc} In the left diagram we illustrate the method of solving
the heat  equation as a boundary value problem for $-\infty<t<\infty$: the solution
(solid line) is defined by the boundary function (a dashed line).
The center  diagram
shows that solution to equation (3.17) is equivalent to
solving the inverse diffusion
equation, that is the classically ill-posed problem. In the right diagram
we show that if the
source $\psi(t)$  (double solid-dashed line) is prepared in a special way,
so that it is a solution to the diffusion equation for an
"early" boundary condition (dashed green line),
we can solve the  problem. In this case the solution (double  green-lilac solid line)
is identical to the solution of the heat equation
with the boundary condition denoted by the  dashed green line.
}\label{Fig:heat}
\end{figure}
Assuming that we know $\psi(t) $ we can write
\be
\phi(t)=\frac{1}{2\pi}\int e^{\lambda\xi^2}\tilde
\psi(\xi)e^{-i\xi t}d\xi
\ee
where $\tilde
\psi(\xi)$ is  the Fourier transform of the source.
 This integral converges for $\lambda>0$ if the function $\tilde
\psi(\xi)$ decreases  fast enough. In particular, we can assume that
there exists $a>0$ such that $\psi(\xi)$ can be presented as
$
\tilde\psi(\xi)= e^{-a\xi^2}\tilde\psi_a(\xi),
$
where $|\psi _a(\xi)|<C$ for $-\infty< \xi<\infty$,
and for $\lambda<a$ the solution to (\ref{mech-exp}) can be presented as
\be
\label{J-sol-infty}
\phi(t)=
{\cal K}[\psi_a](t, a-\lambda),\ee
here $\psi_a$ is the inverse Fourier transform of $\tilde\psi_a(\xi)$,
see Fig.~\ref{Fig:heat}.

For the case $\psi=0$ we get only the trivial solution to equation
(\ref{mech-exp}).

\subsubsection{Exponential equation with quadratic derivatives on a half axis.}
\label{m-2-exp}

Let us  understand the L.H.S. of the equation in (\ref{mech-exp}) as
\be \label{pseudo_app-exp}
  e^{\lambda\partial^2_t}\varphi(t) \equiv
  \frac{1}{2\pi i}\int\limits_{c-i\infty}^{c+i\infty} d\!s\, e^{st}\left[ e^{\lambda s^2}  \tilde{\varphi}(s)
  -  \tilde{r}(s,\lambda)   \right]
  \ee
  where the  residual term $\tilde{r}(s,\lambda)$ is
  $ \tilde{r}(s,\lambda) =\sum_{k=1}^{\infty} \sum_{j=1}^{2k}
  \frac{\lambda^ks^{2k-j}}{k!}d_{j-1},\,\,\,d_{j}=\varphi^{(j)}(0)$.
  
A more general definition with arbitrary constants $d_{j}$ has been discussed in~
\cite{BK:2007}.
This type of definition being applied to an arbitrary function $F(\partial)$
does not guaranty that the factorization form of the operator $F(\partial)$
is the same as the  Weierstrass  factorization of the function $F(z)$.

Performing  summation in the residual term we get the integral
representation in $t$-variable
 \be \label{r-x}
  r(t,\lambda)={\cal J}_{\frac12}[\mathfrak{j}_0](t,\lambda)-
  \frac12\,{\cal J}_{\frac32}[\mathfrak{j}_1](t,\lambda),
\ee
where
  \bea
  {\cal J}_{\frac12}[\mathfrak{j}_0](t,\lambda)&=&\frac{1}{2\sqrt{\pi }}\int\limits_0^\lambda \,d\lambda'
 \frac{e^{-\frac14\frac{t^2}{(\lambda-\lambda')}}}
{\sqrt{\lambda-\lambda'}}\,\mathfrak{j}_0(\lambda'),\,\,\,\mathfrak{j}_0(\lambda)\equiv\frac{\varphi
(\sqrt{\lambda }) - \varphi (-\sqrt{\lambda})}{2\sqrt{\lambda
}},\\
{\cal J}_{\frac32}[\mathfrak{j}_1](t,\lambda)&=&\frac{t}{2\sqrt{\pi }}\int\limits_0^\lambda
\,d\lambda'
\frac{e^{-\frac14\frac{t^2}{(\lambda-\lambda')}}}{(\lambda-\lambda')^{3/2}}\,
\mathfrak{j}_1(\lambda'),\,\,\,
\mathfrak{j}_1(\lambda)\equiv\frac{\varphi (\sqrt{\lambda })+
 \varphi (-\sqrt{\lambda })}{2}. \eea

It seems  that $\mathfrak{j}_0$ and $\mathfrak{j}_1$   contain the values of $\varphi$
for  negative
arguments. But by the construction these values  are calculated via
right derivatives of
function $\varphi$ at zero
$$
\frac{\varphi
(x)+
 \varphi (-x)}{2}=\sum_{n=0} \frac{x^{2n}}{(2n)!}\lim _{y\to +0}\varphi^{(2n)}(y),\,\,
\frac{\varphi
(x)-
 \varphi (-x)}{2}=\sum_{n=0} \frac{x^{2n+1}}{(2n+1)!}\lim _{y\to +0}\varphi^{(2n+1)}(y)
$$
and properly depend on the function $\varphi(t)$ on the half plane
$t\geq 0$.

Taking into account that the  inverse Laplace transform for $t<0$ is equal to zero we
prove  that (\ref{pseudo_app-exp})  is equivalent to
the following representation
 \be
\label{Act} e^{\lambda\partial^2_t}\varphi(t)
= {\cal K}_{+,-}[\varphi](t,\lambda)
+{\cal J}_{\frac32}[\mathfrak{j}_1](t,\lambda)
\ee
where
$
{\cal K}_{+,-}[\phi](t,\lambda)\equiv \int\limits_0^\infty
\frac{e^{-\frac{(t-t')^2}{4\lambda}}-
e^{-\frac{(t+t')^2}{4\lambda}}}{2\sqrt{\lambda\pi} }\,\phi(t')dt'.$
The integrant in the second term in the R.H.S. of
(\ref{Act})
depends only
on even order right derivatives $\lim _{x\to +0}\varphi^{(2n)}(x)$.

Let us remind that  the solution to the  boundary problem
 for the heat equation
\be \label{heq-bc}
\frac{\partial}{\partial \lambda}\Psi(t,\lambda)=
\frac{\partial^2}{\partial t^2}\Psi(t,\lambda),\,\,
\Psi(t,0)=\varphi(t),\,\,
\Psi(0,\lambda)=\mu(\lambda)
\ee
is given by
$
 \Psi(t,\lambda)={\cal K}_{+,-}[\varphi](t,\lambda)+
 {\cal J}_{\frac32}[\mu]((t,\lambda).
$

Representation (\ref{Act}) shows that $\Psi(t,\lambda)=e^{\lambda\partial^2_t}\varphi(t)$
with the exponential operator that we understand  in the sense
(\ref{pseudo_app-exp}) solves
 the heat equation
with the  boundary  condition
$
\Psi(t,0)=\varphi(t)$
and the initial condition
$
\Psi(0,\lambda)=\frac{\varphi
(\sqrt{\lambda })+
 \varphi (-\sqrt{\lambda })}{2}$. This initial condition is  defined by the
 boundary function $\varphi $ on
 the positive half axis.

\begin{figure}[h!]
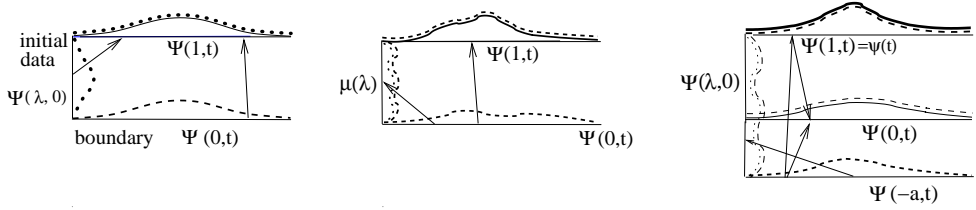

\centering
\includegraphics[width=3.8cm]{bc-ic0-color.eps}$\,\,\,\,\,\,\,$
\includegraphics[width=3.9cm]{bc-ic-def-color.eps}$\,\,\,\,\,\,\,$
\includegraphics[width=3.9cm]{bc-ic-a-color.eps}
\caption{\label{regiona}
The  diagram in the left
 shows solution of the diffusion equation (double blue-red line) with the boundary
 condition at $\lambda=0$
 (dashed blue line),
 $0<t<\infty$ and the initial data $\Psi(0,\lambda)=\mu(t)$ (dotted red line).  The
  diagrams in the center
 shows the solution of the diffusion equation (double solid-dashed blue line) with the special initial data
 (double dotted-dashed blue line),
 defined by the boundary function (dashed blue line). The right diagrams shows solution
 to Eq.(3.17) (double green-lilac solid line)
 for a special source $\psi(t)$ (double solid-dashed line), that is the solution to the diffusion equation for an
 "early" boundary condition (dashed green line) and the corresponding initial function
 (double dotted-dashed green line).
}
\end{figure}

It is instructive to compare boundary conditions for the heat equation on the
axis and on the half axis. Dealing with the heat equation on the axis we fixed
the boundary condition at $\lambda=0$
for all $-\infty<t<\infty$, see Fig.~1, meanwhile on the half axis we fixed
the boundary condition at $\lambda=0$ only for nonnegative values  $0\leq t<\infty$.
The initial condition $\mu(\lambda)$ is defined by the boundary function $\Psi(0,t)$.

Eq.(\ref{mech-exp}) for $0\leq t<\infty$, where we understand
the exponential $e^{\lambda \partial ^2_t}$
 as in (\ref{Act}), can be solved, in particular in the case when  the function $\psi(t)$
 is defined as a solution of the heat equation in $[-a,\lambda]\times [0,\infty]$ with the given boundary function
 $\Psi(-a,t)$ and the initial function  $\mu(\lambda)=\frac12(\Psi(-a,(a+\lambda)^{1/2})+
 \Psi(-a,-(a+\lambda)^{1/2}))$, see the right panel in Fig.2.

\subsubsection{Few comments on nonlinear equations}
Nonlinear Eq.(\ref{EOM_ST0approx_flat}) for $-\infty<t<\infty$,
reduces to the integral nonlinear   equation \be {\cal
K}_{+,-,\xi\,}[\Phi](t,\lambda)= \Phi^3,\ee where $ {\cal
K}_{+,-\,\xi}[\Phi](t,\lambda)=(1+4\xi^2\partial_\lambda){\cal
K}_{+,-\,} [\Phi](t,\lambda) $ and $\lambda=\frac14$.

In the case when one deals  with $t\geq 0$, the boundary condition
due to representation (\ref{Act}) can be presented as an integral
equation with the source
\be
{\cal K}_{+,-,\xi\,}[\Phi](t,\lambda)+{\cal J}_{\frac32,\,\xi}[\mathfrak{j}_1](t,\lambda)= \Phi^3\ee
where
${\cal J}_{\frac32,\,\xi}[\mathfrak{j}_1](t,\lambda)=(1+4\xi^2\partial_\lambda){\cal J}_{\frac32\,}[\mathfrak{j}_1](t,\lambda)
$.
It would be interesting to study solutions of the integral equations with sources.
\section*{Acknowledgements}
It it my pleasure to thank the organizers of ``SFT2010 -- the third
international conference on string field theory and related topics'', held at
the YITP Kyoto,  for hospitality and for a
very  stimulating atmosphere.
I would also like to acknowledge discussions  with  participants of the conference,
and I.V.~ Volovich and   R.V.Gorbachev
 for collaboration and useful discussions.

\end{document}